\documentclass[twoside,twocolumn,english,prl]{revtex4-1}
\usepackage[T1]{fontenc}
\usepackage[latin9]{inputenc}
\usepackage{xcolor}
\usepackage{amsmath}
\usepackage{graphicx}
\usepackage{esint}
\PassOptionsToPackage{normalem}{ulem}
\usepackage{ulem}


\usepackage{babel}
\begin{document}

\title{Record-dynamics description of spin-glass thermoremanent
magnetization: A numerical and analytical study
}

\author{Paolo Sibani$^{1}$ and Jacob M\o ller Kirketerp$^{1}$}

\affiliation{$^{1}$FKF, University of Southern Denmark, Campusvej 55, DK5230,
Odense M, Denmark }

\begin{abstract}
Thermoremanent magnetization data for  the 3D Edwards-Anderson spin glass  are 
generated using the Waiting Time Method as simulational tool
and interpreted using Record Dynamics. We verify that 
 clusters of contiguous spins  are overturned by quakes,
non-equilibrium events linked to record sized energy fluctuations
and show that quaking is a log-Poisson process, i.e.
a Poisson process whose average depends on the logarithm of the system age, counted from
the initial quench.
Our findings compare favourably with experimental thermoremanent magnetization findings
and with the spontaneous fluctuation dynamics of the E-A model.
 The logarithmic growth of the size of overturned clusters
 is related to similar experimental results and to the growing length scale
of the spin-spin spatial correlation function.
 The analysis buttresses the applicability of the Waiting Time Method as a simulational tool
 and of Record Dynamics as coarse-graining method 
  for aging dynamics.
\end{abstract}
\maketitle

\section{Introduction}
The multi-scale relaxation 
process  called \emph{aging} is  observed in
  e.g.
spin-glasses~\cite{Vincent96,Nordblad97,Vincent07}, colloidal
suspensions~\cite{Hunter12,Sibani19}, vortices in superconductors~\cite{Nicodemi01} and
 evolving biological and cultural ecologies~\cite{Becker14,Nicholson16,Arthur17}. Important  aspects of aging phenomenology  have been elucidated by
spin glass linear response experiments~\cite{Vincent96,Nordblad97,Vincent07}, including
thermo-remanent magnetization (TRM) studies of
memory and rejuvenation effects~\cite{Alba86,Zotev03,Vincent91,Vincent96},
sub-aging and end of aging behavior~\cite{Rodriguez03,Kenning06}.
 Numerical simulations~\cite{Rieger93,Kisker96,Sibani07,Belletti09,BaityJesi17,Sibani18} of the Edwards-Anderson (E-A) model~\cite{Edwards75}
 provide additional  insight 
and a test  of  theoretical assumptions. 

In TRM experiments, the system  is  thermalized  in a magnetic field $H$ and then  
thermally quenched   at time $t=0$~\footnote{We gloss here over the fact that an instantaneous quench
 is not experimentally achievable.}  below
 $T_c$, the spin-glass critical temperature.  At $t=t_{\rm w}$ the field  is cut
and the magnetization decay is measured for $t>t_{\rm w}$.
 Note that   field removal and  
 time origin are  usually taken to coincide, in which case 
 $t$  notationwise corresponds to  our  $t-t_{\rm w}$.

Record Dynamics~\cite{Anderson04,Sibani06} (RD)  deals with 
 complex systems~\cite{Sibani13} lacking time translational
invariance and evolving through a series of  equilibrium-like configurations
of increasing duration.
 This experimental and/or observational background can be theoretically associated to
 a hierarchy of dynamical barriers and, in a second step,
to  a hierarchy of nested ergodic components~\cite{Palmer82}, each 
predominantly found  in a stationary, or pseudo-equilibrium, state.

RD highlights   the irreversible events,  called \emph{quakes}, bringing  the 
system from one pseudo-equilibrium state to the next. It posits that quakes constitute 
a Poisson process whose average depends on the logarithm of time, for short 
a log-Poisson process
The transformation $t\rightarrow \ln t$ then
produces a log-time homogeneous  coarse grained description of 
aging, yielding  specific predictions for experimental and  numerical observations.

An  analysis of experimental TRM data~\cite{Sibani06a} and   simulations of
the  zero field cooled magnetic linear response
of the 3D E-A model with Gaussian couplings~\cite{Sibani07}
both  make use of RD, and  
identify quakes as anomalously large magnetic fluctuations.
 More recently~\cite{Sibani18}, the same model was simulated 
 for a range of low temperatures in zero field, with the Waiting Time Method  (WTM)~\cite{Dall01} 
 as simulational tool.
  Quakes are associated to records   in the time series of energy
 values produced by the simulation,  and  real valued event  times are assigned  to them,
 providing the statistics needed to ascertain the  log-Poissonian nature of the quaking  process.
 
   Following the same methodology,   the present  study aims to show 
 the agreement between WTM simulations  and experimental descriptions, and  the ability of RD to 
 predict the  key features of spin-glass  dynamics.
 To this end, we first demonstrate the log-Poisson nature of
the quaking process, and then check RD predictions
 on  the time dependence of the TRM and the system excess energy~\cite{Sibani06a,Sibani07}, Finally, the real space effect   of quakes is described in terms of
the near simultaneous overturning  of 
 a cluster of adjacent spins, i.e. a spin flip cascade over a barrier. This is  compared with experimental results,
 where   clusters of similar nature
 are extracted from TRM traces~\cite{Joh99}.

The rest of the paper is organized as follows:
In Section~\ref{MM} the model and the simulation method is briefly described.
Section~\ref{RR} is devoted to the simulation results
 and
Section~\ref{SC} to a summary and a conclusion.

 \section{Model and Simulation Protocol}
\label{MM}
This section closely follows Ref.~\cite{Sibani18} to which we refer for additional details.
   Essential information, including differences  from the above  reference 
 are given here for the reader's convenience.
 
We consider an Ising E-A spin glass~\cite{Edwards75}
placed on a cubic grid with linear size $L=20$ and periodic boundary
conditions. Each of the $2^{N}$ configurations is specified by the
value of $N=L^{3}$ dichotomic spins, and has, in a magnetic field $H$,
an energy given by 
\begin{equation}
{\mathcal H}(\sigma_{1},\sigma_{2},\ldots\sigma_{N})=
-\frac{1}{2}\sum_{i=1}^{N}\sum_{j\in{\mathcal{N}}(i)}J_{ij}\sigma_{i}\sigma_{j} -H\sum_{i=1}^{N}\sigma_{i},
\label{En_def}
\end{equation}
where $\sigma_{i}=\pm1$ and where ${\mathcal{N}}(i)$ denotes the
six nearest neighbors of spin $i$. For $j<i$, the $J_{ij}$s are
drawn independently from a Gaussian distribution with zero average
and unit variance. Finally, $J_{ij}=J_{ji}$ and $J_{ii}=0$. All
parameters are treated as dimensionless. 
For $H=0$, the model has a phase transition from a paramagnetic
to a spin-glass phase at a critical temperature which in Ref.~\cite{Katzgraber06} is 
estimated to be  $T_{\rm c}=0.9508$. 

Our  system is thermalized at temperature $T_{0}=3$ in a  magnetic  field $H=0.1$, 
  instantaneously
quenched at time $t=0$ to temperatures $T=0.5, 0.6 \ldots 1$
and then left to age isothermally. At  time $t_{\rm w}$ the magnetic field is removed 
and the magnetization decay 
is  observed  as a function of  time. We consider three values of $t_{\rm w}$,
$t_{\rm w}=10, 100$ and $200$.

 The real space manifestation of the quakes in the E-A model  are cascade events, where clusters
 of adjacent spins flip coherently.
 The distinction  between cascade events  and scattered flips is moot in standard MC algorithms, e.g. 
parallel tempering~\cite{Katzgraber03}, since  consecutive  queries are always  spatially uncorrelated
and  the shortest available `time' scale is a MC sweep. 

The rejectionless 
Waiting Time method (WTM)~\cite{Dall01}, where `Waiting Time' refers to
the time between two successive moves,
 generates a
 Markov chain  closer to   a physical relaxation process  than
 is the case for  standard MC algorithms.

Each basic degree of freedom, e.g. a spin $i$, performs a Poisson process
whose  characteristic time scale $\tau_i$  depends on  the interactions with its neighbors.
Any state change of the neighbors  resets the  process and requires the 
 recalculation of  $\tau_i$.
 Every spin has a flip-time at which it would flip if nothing else happened
  and the spin  actually flipped is the one with the earliest  flip-time.
  Each  spin flip in a simulation is thus associated with 
an intrinsic real-valued 
time variable $t$ and spatially and temporally localized dynamical  events 
are possible and can be precisely identified.

 When  the WTM is applied to the E-A model, spin $i$   stays put
for an exponentially distributed time interval $\tau_i$, unless one of 
its neighbors flips.
 The  mean waiting time  $\tau_i$ to its next possible move,
 hinges on the  energy change $\Delta E_i$ such move would entail.
 Assuming $\Delta E_i>0$, the situation is locally 
 metastable, but an updated  value of $\Delta E_i$ due to activity in the
 neighborhood requires a re-calculation of the waiting time. 
  In the unstable situation where $\Delta E_i<0$, the waiting time is with high probability very short
  and the spin quickly flips. A flip can in turn create a new unstable situation in the neighborhood,
  and
  iterating this process   generates a sequence of  
neighboring spins quickly flipping  one  after the   other.  When 
no further energy loss is possible the process stops  and a new metastable configuration is created 
differing from its predecessor in the orientation of a spatially connected  cluster of spins.

To detect a quake   we follow \cite{Sibani18},
 and use   two `record' energy values 
 in combination with  a subdivision of the time axis in short intervals of 
 equal  duration $\delta t$. The quantity $E(t)$ is the energy as function  of time.
 The two record values used  are the `best so far'  energy $E_{\rm bsf}(t)$
and the `highest so far' energy $E_{\rm h}(t)$. 
The former  is the least energy seen during the simulation up to the `current' time $t$,
\begin{equation}
E_{\rm bsf}(t)=\min_{0<t'<t}(E(t'))
\end{equation}
 and the latter  
is the largest energy value seen, relative to the best so far energy,
\begin{equation}
E_{\rm h}(t)=\max_{0<t'<t}(E(t')-E_{\rm bsf}(t'))
\end{equation}

For simplicity, the system energy $E$ is
measured relative to $E_{\rm bsf}$ and its current position on the time axis
is continuously tracked. A
quake alert device  with three  states $0,1$ and $2$  is utilized
 for quake   identification.
State $0$
covers standard fluctuation dynamics,
state $1$ is reached when   $E_{\rm h}$ is updated, i.e. increased, and 
state $2$ when  $E_{\rm bsf}$ is subsequently updated, i.e. decreased.
At this point an unfolding quake is detected and  the alert level is reset to $0$.
The quake event  is deemed to have terminated once time exceeds
the boundary of the current $\delta t$ sub-interval  of the time  axis.
The spin cluster which changed orientation during the quake is 
identified, and the time at which the quake occurred is registered.
In   \cite{Sibani18}, the same procedure is followed  except that the detection device 
has there two states rather than three. State $1$, which  triggers quake detection,
is reached if either $E_{\rm h}$ or $E_{\rm bsf}$ is updated. We modified the algorithm 
to avoid an excessive number of events being registered right
after  magnetic field removal.  

\section{Results}
\label{RR}
 \subsection{Log-Poisson statistics}
\label{LP}
In this section 
we show that the quakes extracted from our TRM data,
i.e. after field removal at times  $ t_1 < t_2 \ldots <t_k\ldots $,
with $t_1>  t_{\rm w} $,
are a Poisson process whose  average 
depends on the logarithm of $t_k/t_{k-1}$. Since the analysis deals
with the distribution of inter-quake times, the waiting time $t_{\rm w}$
does not explicitly enter the discussion.

A quake that flips a cluster 
can facilitate the overturning of 
a partly overlapping or  neighboring cluster.
Quakes can therefore  be   interdependent  in regions of configuration space  
extending well beyond the correlation length associated to thermal equilibrium 
 fluctuations.
\begin{figure}[t]
\hfill{}\includegraphics[bb=20bp 180bp 560bp 600bp,clip,width=1\columnwidth]{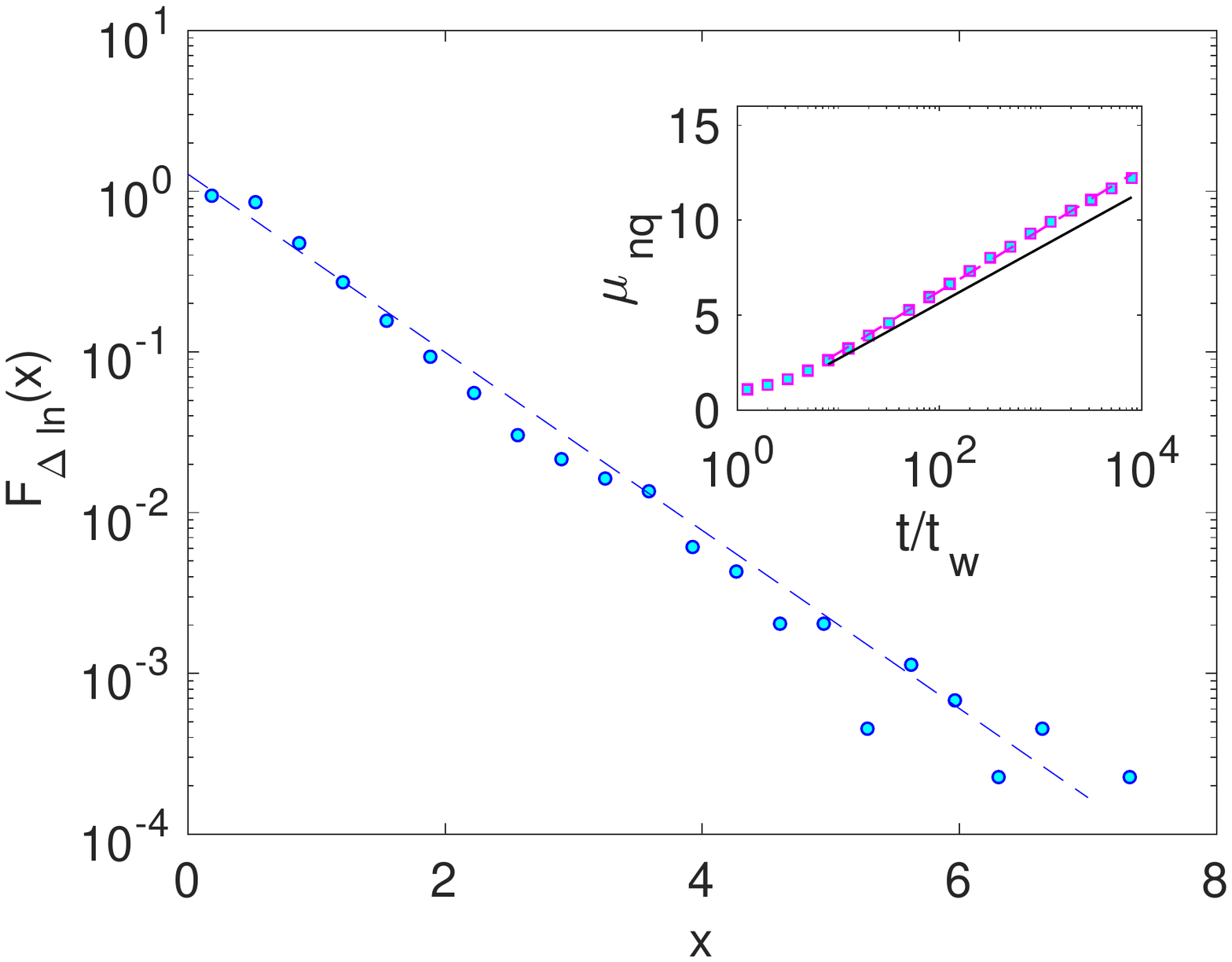}\hfill{}
\caption{Symbols: PDF of  `logarithmic waiting times' $\Delta{\rm ln}$, 
for the aging temperatures $T=.5$.
Dotted line: fit to the exponential form $y(x)=r_{\rm q}e^{-r_{\rm q}x}$ with $r_{\rm q}=1.275$. 
Insert: the squares show  the average number of quakes  vs the logarithm of $t/t_{\rm w}$  for $t_{\rm w}=10$.
The
hatched line depicts the fitted function $y=r'_{\rm q}\ln t -0.2388$ with $r'_{\rm q}=1.410$
while  the full line makes use of the logarithmic rate $r_{\rm q}$.
That $r'_{\rm q}\ne r_{\rm q}$ shows a discrepancy between the two estimates of the logarithmic rate of events.
 }   
\label{fig:DLTS} 
\end{figure}

In a  large systems, quake to quake correlations, not to be confused with thermal correlations,
 will eventually die out and temporally close but spatially distant successive quakes 
will then be uncorrelated.
We focus on  a spatial   domain where quakes are all interdependent, and where  the transformation $t \rightarrow \ln t$
captures all  temporal correlations. Quaking 
 is  then   described by a memoryless Poisson process 
whose average is proportional to the logarithm of time, for short a log-Poisson process.

To ascertain if  this is actually the case, it suffices to check whether the 
log-waiting times between successive events have an exponential PDF with unit average.
Log-waiting times are simply defined in terms of the occurrence time $t_k$ of the $k$'th quake as
$\tau_k=\ln(t_k/t_{k-1})$. Their empirical PDF is predicted to have the form 
\begin{equation}
F_{\rm \Delta ln}(x)=r_{\rm q}e^{-r_{\rm q} x},
\label{FDelta ln}
\end{equation}
where the constant $r_{\rm q}$, the logarithmic quaking rate, is unity. As shown in the
main panel of  Fig.\ref{fig:DLTS},
the exponential PDF fits our data, though with  a  value of $r_{\rm q}$ somewhat higher
than predicted. The insert of the same figure shows that the number of quakes that  fall in the interval $(0,t]$, averaged
over all trajectories,  grows as $\mu_{\rm  nq}(t)=r'_{\rm q}\ln(t)$.
The logarithmic growth is as predicted by RD,
but  $r'_{\rm q}>r_{\rm q}$ while
 $r_{\rm q}=r'_{\rm q}$ according to theory.
 The two lines in the insert of Fig.\ref{fig:DLTS}
  highlight the difference.

To conclude  this section, briefly consider the situation where \eqref{FDelta ln} does not fit the
 empirical distribution of log waiting times. 
Plainly, the discrepancy can arise if RD does not apply to the problem at hand.
The other possibility is that data are collected over a spatial domain large enough
to accommodate uncorrelated quakes. In this case the average number of 
quakes will still grow logarithmically, with a pre-factor reflecting the number of 
uncorrelated domains contained in the system. Since the waiting time between uncorrelated events is 
exponentially distributed,  once  uncorrelated quakes dominate the PDF of
waiting times ---rather than log-waiting times--- between successive quakes will be exponential.
To recover Fig.~\ref{fig:DLTS} one needs to  consider domains of reduced size. Ref.~\cite{Sibani19}
gives an example where this situation arises. Alternatively, one can follow Ref.~\cite{Boettcher18} 
and note   that uncorrelated quakes produce a peak
in $F_{\rm \Delta ln}(x)$ for small waiting times near $x=0$. For sufficiently large $x$, the decay remains unchanged, i.e.
exponential. 

\subsection{Macroscopic data}
\label{Macro}
\begin{figure}[t]
\hfill{}\includegraphics[bb=20bp 180bp 560bp 600bp,clip,width=1\columnwidth]{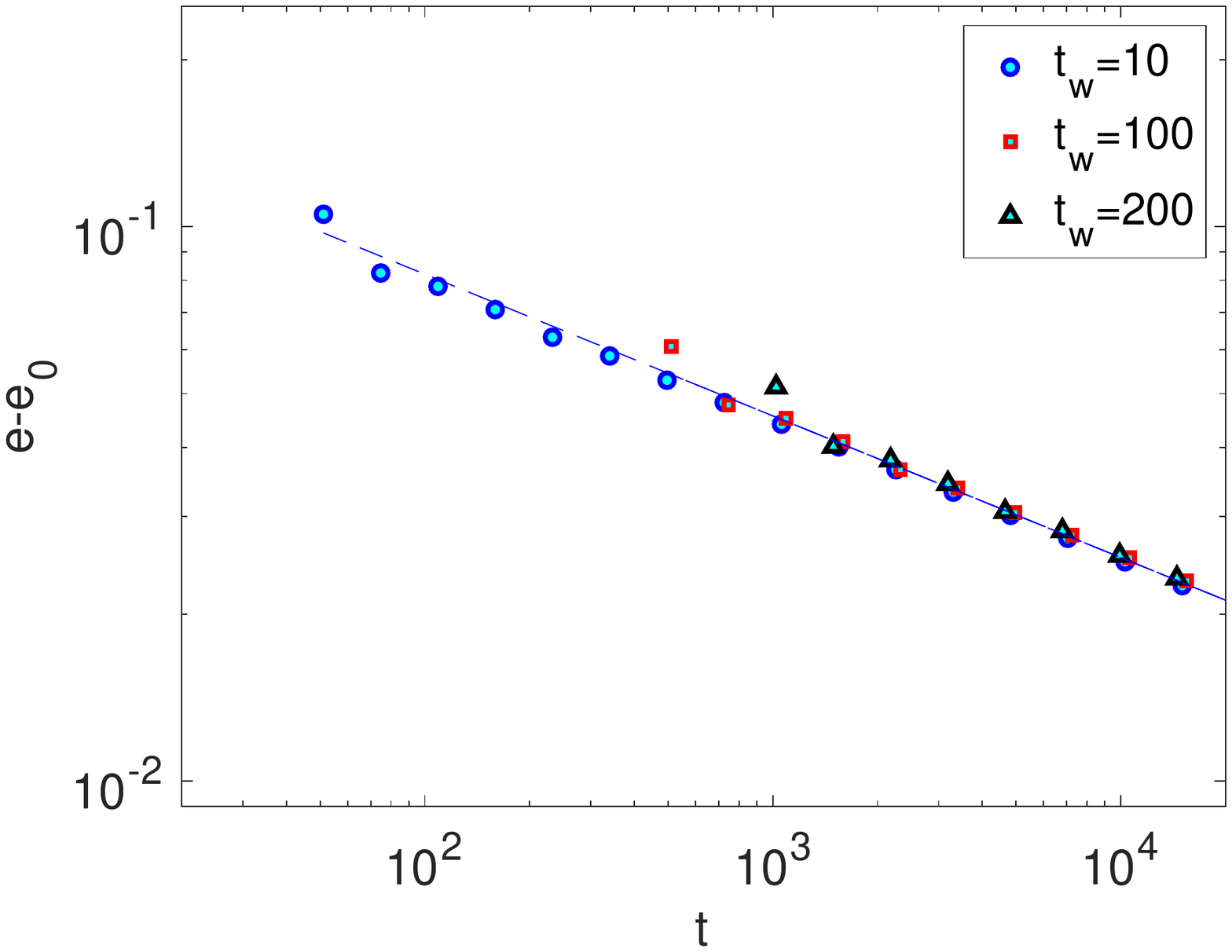}\hfill{}
\caption{Symbols: the energy per spin, with a fitted ground state energy value subtracted, is plotted vs. time
for three  systems with different field removal times.
The dotted  line is a fit to the  power law $y(t)=a t^{-\alpha}$, where $a$ and $\alpha$ are free parameters.
 }    
\label{fig:EvsT} 
\end{figure}

Unlike experiments,  numerical simulations provide easy access
to the system energy.
 Figure~\ref{fig:EvsT} shows the difference
 $e(t)-e_0$ between the energy per spin and its ground state value
 plotted vs. time $t$. Three data sets are included, corresponding
 to different values of $t_{\rm w}$, all three fitted to the same the power law 
\begin{equation}
e(t)-e_0=a t^{\lambda_e},
\label{energyfit}
\end{equation}
 where $e_0=-1.6813,\; a=0.2664 $ and $\lambda_e=-0.2557 $ are  fitting parameters.
  As expected, in a linear response experiment,  no significant energy dependence 
  appears on the field removal  time   $ t_{\rm w}$.
  
Equation~\eqref{energyfit} was proposed in~\cite{Sibani90a}  to
estimate the ground state energy $e_0$ as the value producing the power law
decay. The  decay was  observed in isothermal simulations of the E-A model, using the WTM~\cite{Sibani07}   and parallel tempering~\cite{Belletti09}.  
See Eq.~(31) and Table~5 of the latter  reference for the  correspondence to the present notation.
These authors tentatively attribute the power law decay   to the system being critical for a range of temperatures.
 Our estimate $e_0\approx -1.68$ is nearly identical to that of \cite{Sibani07}, while
  ref.~\cite{Belletti09}, where much larger systems are considered with two-valued couplings $J_{i,j}=\pm 1$
   finds values close to $-1.77$.
 The exponent $\lambda_e$ is a negative and linearly decreasing function of temperature, with the value  $\lambda_e(T=0.5)\approx -0.25$
 from~\cite{Sibani07} close to our current estimate.
 Ref.~\cite{Belletti09} finds  $\lambda_e(T=0.6)= -0.193$
  somewhat higher  than our  $T=0.5$ value. The mismatch is related to
 algorithmic details. Note however that \cite{Sibani07} and \cite{Belletti09} agree on $\lambda_e$ being a decreasing function of $T$.

 Our main interests lies  not in how to best estimate the ground state energy, but in the fact that
RD predicts  the power law decay, in a way   unrelated to critical behavior.
Assuming that only quakes can lower the energy,  $e(n)$ is a function of the number $n$
of quakes occurring in the interval $(t_{\rm w},t)$. Furthermore,
each quake can be expected to  decrease the energy
difference \mbox{$\Delta_e(n)=e(n)-e_0$}
 by a constant fraction. 

Our assumption entails
\begin{equation}
\Delta_e(n)=\Delta_e(0)x^n,
\end{equation}
$x$, with $0<x<1$. 
In order to extract a time dependence, the  expression  must be averaged 
 over the  Poisson distribution of the number of quakes falling in the observation interval
$(t_{\rm w},t)$ .

Taking $\Delta_e(t)=e(t)-e_0$, this   yields
\begin{eqnarray}
\Delta_e(t)&=&\Delta_e(t_{\rm w}) \left( 	\frac{t}{t_{\rm w}} \right)^{-r_{\rm q}} \sum_{n=0}^\infty \frac{(x r_{\rm q} \ln(t/t_{\rm w}) )^n}{n!} \nonumber \\
&=&
\Delta_e(t_{\rm w})\left( \frac{t}{t_{\rm w}} \right)^{-r_{\rm q}(1-x)}.
\end{eqnarray}
In an RD description,  the exponent $\lambda_e$ characterizing the energy decay  is given 
by $\lambda_e=-r_{\rm q}(1-x)$, which is unrelated to any critical exponent. 
Power law behavior comes indeed naturally 

in processes involving activation  over a \emph{hierarchy} of barriers, see~\cite{Uhlig95}and references therein. 

Turning now to the Thermoremanent Magnetzation (TRM), we use
 the gauge transformation $\sigma_{i}\rightarrow\sigma_{i}(t_{{\rm w}})\sigma_{i},\;J_{ij}\rightarrow\sigma_{i}(t_{{\rm w}})\sigma_{j}(t_{{\rm w}})J_{ij}$
to  map it  into the correlation function
\begin{equation}
C(t_{{\rm w}},t)=\sum_{i}\langle\sigma_i(t_{{\rm w}})\sigma_i(t)\rangle.
\label{corr_f}
\end{equation}
Modulo multiplicative constants, the two functions hold   equivalent
information and since the general form of the autocorrelation function  is theoretically available,
we can use it to fit the TRM.   

We consider  the de-correlation induced by  the quakes, which allow the system to
equilibrate in increasingly large ergodic components.
Quaking is log-time homogeneous stochastic process, involving
a set of interacting mesoscopic dichotomic variables, our clusters.
Even though a  formal description of how clusters interact is lacking,
general arguments~\cite{Sibani13} lead to
\begin{equation}
C(t_{{\rm w}},t)=\sum_{i}     w_i \exp(\lambda_i \ln(t/t_{\rm w})),
\label{corr_f2}
\end{equation}
where the  $\lambda_i $s  are negative eigenvalues associated to  the normal modes of the relaxation process
and  all $w_i$ are positive real numbers.

\begin{figure}[t]
\hfill{}\includegraphics[bb=20bp 180bp 560bp 600bp,clip,width=1\columnwidth]{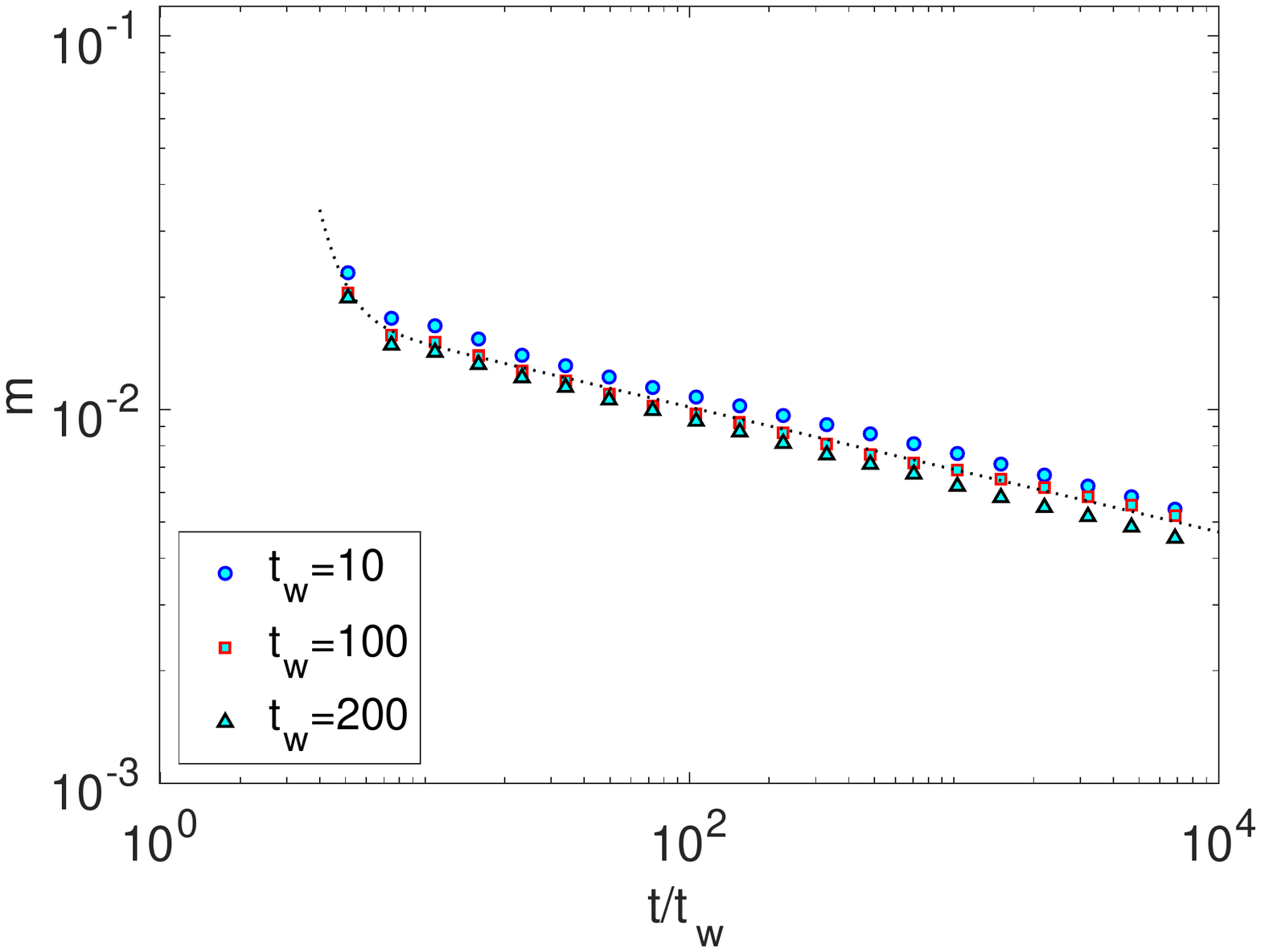}\hfill{}
\caption{Symbols: Thermoremanent magnetization data for $T=.5$  plotted vs  scaled time $t/t_{\rm w}$.
Three data sets are shown corresponding to  the $t_{\rm w}$ values given in the legend.
Dotted line: fit using two terms of the expansion~\eqref{corr_f2}.
 }   
\label{fig:TRM} 
\end{figure}

\begin{figure}[t]
\hfill{}\includegraphics[bb=20bp 180bp 560bp 600bp,clip,width=1\columnwidth]{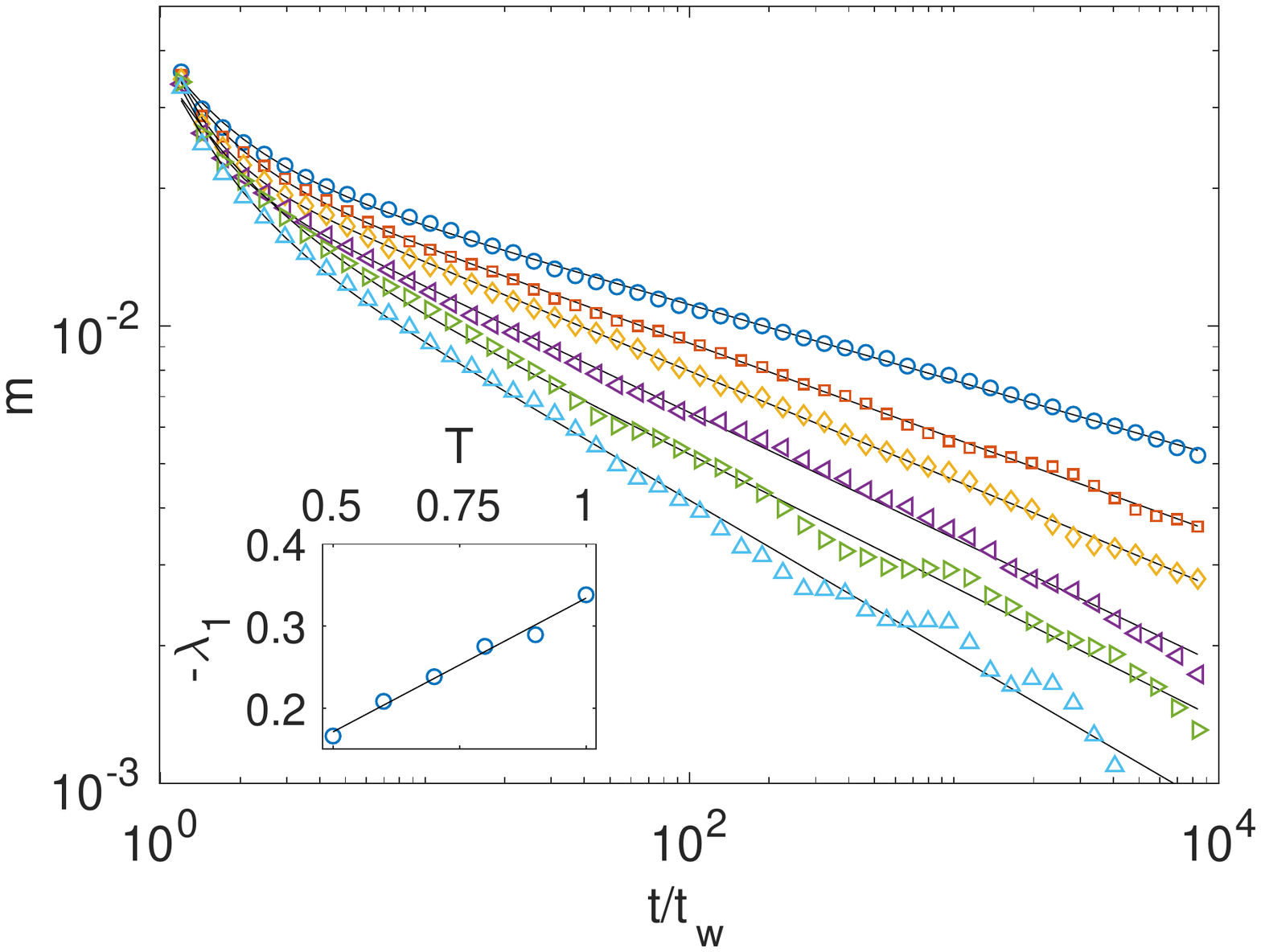}\hfill{}
\caption{Symbols: Thermoremanent magnetization data for isothermal aging at  temperature $T$. From top to
bottom,  data sets  collected at 
 $T=0.5,0.6,0.7,0.8, 0.9$ and $1$  are plotted vs  the scaled time $t/t_{\rm w}$, where
 $t_{\rm w}=10$.
Dotted lines: fits using two terms of the expansion~\eqref{corr_f2}. Insert: the exponent of the dominant term in the expansion
is plotted vs the temperature $T$. The line is the linear fit $-\lambda_1=0.33 T$.
 }   
\label{fig:TRM5T} 
\end{figure}

 TRM time series are plotted   in fig.~\ref{fig:TRM} as symbols 
   vs the scaled time $t/t_{\rm w} $. Neglecting  small  deviations from
pure aging~\cite{Alba86},
 all data are fitted by the same function, represented  by a staggered line 
 and obtained by 
 truncating  Eq.~\eqref{corr_f2} to two terms, each having the form $w_i t^{\lambda_i \frac{t}{t_{\rm w}}}$.
 The fitted exponents are 
  \mbox{$\lambda_1=-0.167$} and  \mbox{$\lambda_2=-5.418$} 
with   pre-factors $w_1=0.022$ and $w_2=30.83$.
The first term is well approximated by a logarithm for the range of the abscissa, while the second only matters for 
values of the latter close to one.
Note that power-laws eventually vanish. This  ensures that
TRM traces with different $t_{\rm w}$ values approach both zero and each other, when plotted as a function 
of the observation time $t_{\rm obs}=t-t_{\rm w}$.  This feature  has been measured experimentally~\cite{Kenning06},
were it was termed  `end of aging'.
Figure~\ref{fig:TRM5T} shows six TRM traces, taken  for $t_{\rm w}=10$ at. temperatures $T=0.5,0.6\ldots1$,
with the staggered line depicting fits based, as before, on Eq.~\eqref{corr_f2}.
The exponent closest to zero is plotted in the insert vs the temperature $T$, on which it has  the linear dependence
shown by the line $-\lambda_1=0.33 T$.
.
Finally note that our earlier  RD analysis of TRM experiments~\cite{Sibani06} is also based on  Eq.~\eqref {corr_f2},
but  utilizes three rather than two of its terms. The dominant exponent  is there 
closer to zero and almost temperature independent, which produces a near logarithmic TRM decay.

\subsection{Mesoscopic  real space properties}
\label{RS}The size of thermally correlated domains 
 has attracted both numerical~\cite{Rieger93,Kisker96,Sibani94a,Andersson96,Belletti09,BaityJesi17}
and experimental~\cite{Joh99,Wood10} attention.
As the aging process surmounts increasingly high free energy  barriers and thermal equilibrium is reached in 
increasingly larger ergodic components, the domain size is expected to grow. The  process has been 
followed by measuring  the  correlation length  $\xi(t,T)$ associated to the spin-spin correlation function~\cite{Rieger93,Kisker96},
by identifying  spatial domains using  projections on an alleged ground state obtained by 
annealing~\cite{Sibani94a,Andersson96}, or, experimentally, by an
 analysis of  TRM data introduced by Joh et al.~\cite{Joh99}.
The method   follows $S(t_{\rm obs})$,
the derivative of the TRM with respect to the logarithm of the observation time, in our notation $t_{\rm obs}=t-t_{\rm w}$,
and in particular, the position of its maximum, which 
 demarcates the cross-over at $t_{\rm obs}=t_{\rm w}$ from 
pseudo-equilibrium to non-equilibrium dynamics.
Increasing  the applied field reduces the corresponding  free energy barriers, and thereby 
  shifts  the maximum of $S(t_{\rm obs})$  to the left, 
by an amount  proportional  to the size of  the spin clusters participating  in the barrier crossing process.

These  clusters are observed directly in numerical simulations,  here and in~\cite{Sibani18}, as the coherent movement of 
adjacent spins triggered by a quake.
To  obtain the size of clusters flipped `near' a certain  time $t$,
the simulation time is subdivided into $41$ bins of equal logarithmic
duration. Choosing $t$ at the boundary between two bins, all clusters
overturned at times within these bins are assigned to  $t$.

\begin{figure}
\vspace{-3cm}
\includegraphics[width=\linewidth]{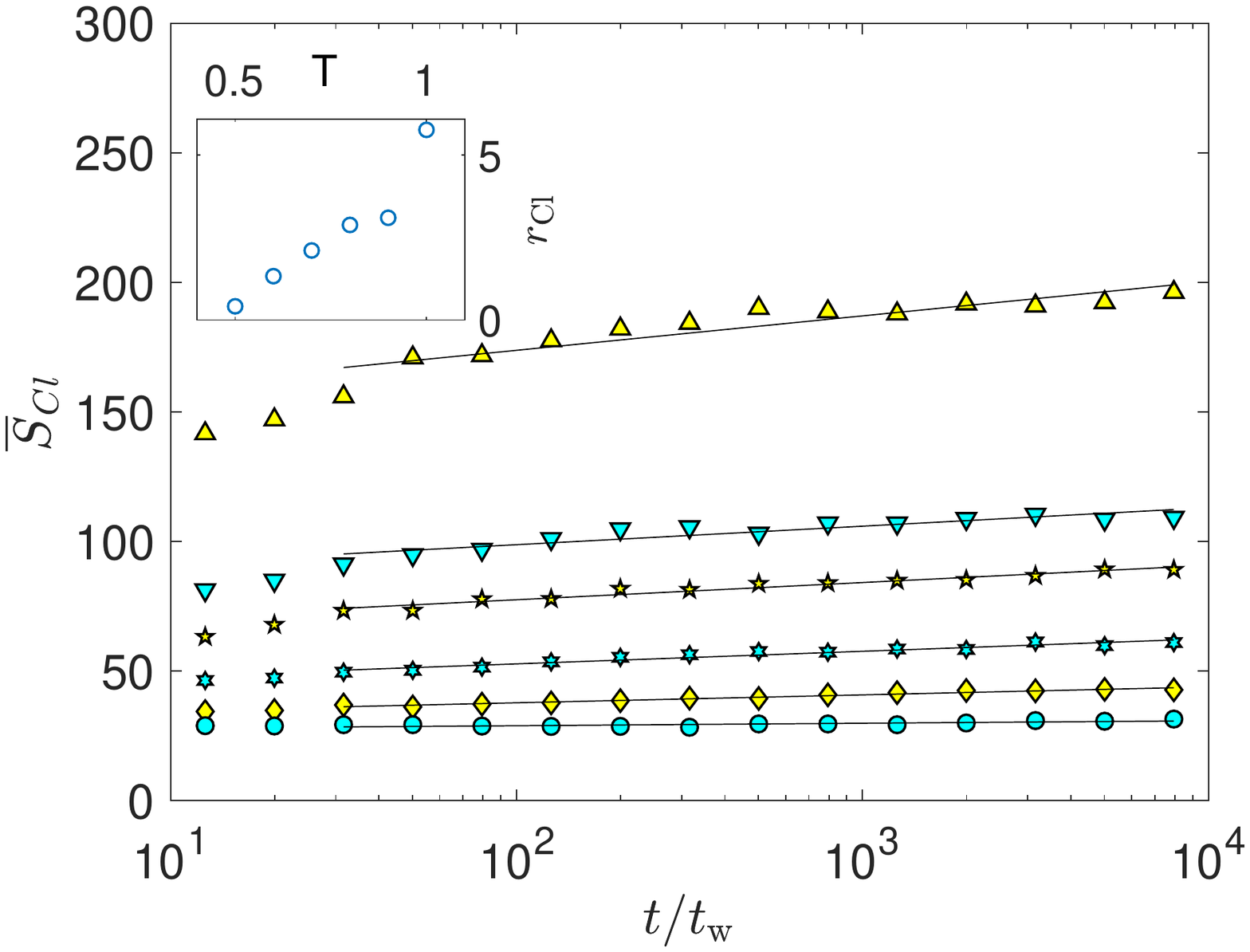}
\vspace{-3cm}
\caption{ Symbols: the average cluster size  ${\overline S}_{Cl}$
is plotted vs the scaled
time $t/t_{\rm w}$ for different temperatures $T$ and $t_{\rm w}=10$.  From top to bottom,
 the data sets  are collected at $T= 1, 0.9, 0.8, 0.7, 0.6$ and $0.5$.
The lines are linear fits of ${\overline S}_{Cl}$ to $\log t/t_{\rm w}$.
 The insert  depicts the corresponding logarithmic rates vs
 the temperature $T$.
}
\label{fig:clusters}
\end{figure}

Figure \ref{fig:clusters} depicts the average size ${\overline S}_{Cl}$ of clusters  flipped near time $t$ vs the scaled time variable $t/t_{\rm w}$
for $t_{ \rm w}=10$ and temperatures, from top to bottom, $T=1, 0.9, 0.8, 0.7, 0.6$ and $0.5$. The lines are fits showing the logarithmic growth
of the cluster size, and the corresponding  rates are  plotted  in the insert vs the  temperature 
 $T$, 
 The statistics is obtained using 1000
independent simulations for each parameter value. The data can be fitted by the expression
\begin{equation}
S_{\rm Cl}(t,T)=(aT+b) \ln(t/t_{\rm w}) +C,
\end{equation}
where $a=9.3937$ and   $b= -4.44569$ and $C$ is a constant.
Clearly, the logarithmic growth  rate should never be negative, and our linear fit is inadequate 
for temperatures below $T=0.5$.

To connect the cluster size to the correlation length $\xi(t,T)$
requires 
theoretical assumptions~\cite{Joh99,BaityJesi17}
which however seem hard to verify unequivocally.
The difficulty arises
because correlation lengths of all origins,
even when  observed  for long time intervals, only have   a modest variation, 
 typically spanning  less than a decade.  
 Furthermore, a simple dimensional connection between correlation length and the average size of flipped clusters 
could be strongly affected by  the
spatial heterogeneity, since all spins participate 
in reversible thermal fluctuations, but many  are not involved in quakes at all.

Ref.\cite{Kisker96} shows that  both  a power law and a logarithm
 can fit 
the time dependence of the correlation length  $\xi(t,T)$ for the  E-A model of with Gaussian bonds,
 and \cite{Wood10} collects and discusses results from many different sources,
all fitted using two power laws, with small exponents linearly dependent on the ratio
of the temperature to its critical value. 

The time dependence of the cluster sizes extracted from experimental data  are shown in~\cite{Joh99}, 
on a linear scale vs a logarithmic time scale
in their Fig. 4 and on a log-log scale in their Fig.5. Assuming that the characteristic cluster size is the third power of the correlation length 
$S_{\rm Cl}(t,T)\propto \xi(t,T)^3$, the 
experimental data  can be fitted using, for the correlation length,  a power law with a small exponent
with a  linear $T$ dependence, or activated dynamics, i.e a logarithm elevated to a power of order five.
Baity-Jesi et al.~\cite{BaityJesi17} used the same type of analysis as~\cite{Joh99} to obtain the cluster size from  large scale simulations
of the  $J= \pm 1$ E-A model. They also express the cluster size in terms of a correlation length elevated to a power.

To conclude, our average cluster size features a slow systematic increase with time, that can be fitted by both  a logarithm and  a power law, in broad agreement
with previous findings. The  correlation length, on which we do not have direct results, is known  to have a
qualitatively similar growth. The precise connection between cluster size and correlation length needs, we believe,  further numerical verification.

\section{Summary and Conclusions}
\label{SC}

Spin glass phenomenology is experimentally well
described~\cite{Nordblad97,Vincent07}, while much  theoretical understanding relies on
partly competing  approaches, conceptually rooted 
in equilibrium statistical mechanics.
The ambition of Record Dynamics (RD) is to offer a simple and  general
coarse-graining method  to analyze the dynamics of a class of  metastable systems, to which spin glasses 
belong.

 In this work, log-time homogeneity and
 the ensuing  RD predictions 
for the decay of the excess energy  and the  TRM 
are verified in a spin glass thermoremanent magnetization numerical simulation.

In thermal equilibrium,  spontaneous fluctuations and linear response    convey the same dynamical
information, since the initial state of a linear response experiment could also have arisen from a spontaneous fluctuation.
The situation is more complicated in  a  TRM  setting, because the barrier structure controlling the dynamics
depends on the external magnetic  field~\cite{Joh99,Guchhait17}.
Barrier climbing has a temperature dependence, described
by a  temperature scaling exponent. The latter is
  $\alpha=1$ for the TRM data, as expected from quasi-equilibrium fluctuations,
  but  $\alpha=1.75$ for spontaneous fluctuation, a value explained in \cite{Sibani18}
  in terms of the density of states near local energy minima.
 Apart from this difference,
   the present results concur  with Ref.~\cite{Sibani18}.
This, combined with the agreement  with both experimental and other numerical results,
confirms the validity of the WTM  as simulational tool.

  Finally, we show that the size of spin clusters overturned by quakes grows logarithmically in time, 
  in agreement with~\cite{Sibani18}. Wood~\cite{Wood10} collects correlation length data of different origins, and
  shows that they all  can  be fitted by a power-law with a small exponent, linearly dependent on the temperature.
  We doubt that the correlation length and the  cluster size have  a simple geometric relation. Furthermore,
   considering that  the correlation length typically only varies over less than a decade, the difference between 
  a power law and a logarithm is moot, and the correlation length could well grow logarithmically in time.
  
  To conclude, together with Ref.~\cite{Sibani18} this work buttresses a  RD description of complex dynamics, and 
 confirms  that the  WTM algorithm, on which our data analysis relies, generates a Markov chain in configuration
 space which  closely mimics  the dynamics of  experimental systems.
\begin{acknowledgments}
P. Sibani thanks  Stefan Boettcher for interesting conversations on  Record Dynamics and for his comments on this work.
\end{acknowledgments}


\begin{thebibliography}{10}
\bibitem{Vincent96}
{Eric Vincent, Jacques Hammann, Miguel Ocio, Jean-Philippe Bouchaud, and
  Leticia F. Cugliandolo}.
\newblock Slow dynamics and aging in spin-glasses.
\newblock {\em SPEC-SACLAY-96/048}, 1996.

\bibitem{Nordblad97}
Per Nordblad and Peter Svedlindh.
\newblock {Experiments on spin glasses}.
\newblock In A.~P. Young, editor, {\em {Spin Glasses and Random Fields}},
  page~1. World Scientific, 1997.

\bibitem{Vincent07}
 {Eric Vincent},
\newblock Ageing, rejuvenation and memory: The example of spin glasses.
\newblock {\em Lecture Notes in Physics}, 716: 7--60, 2007.

\bibitem{Hunter12}
Gary~L. Hunter and Eric~R. Weeks.
\newblock The physics of the colloidal glass transition.
\newblock {\em Rep. Prog. Phys.}, 75:066501, 2012.

\bibitem{Sibani19}
Paolo Sibani and Carsten Svaneborg.
\newblock Dynamics of dense hard sphere colloidal systems: a numerical
  analysis.
\newblock {\em Phys. Rev. E}, 99:042607, 2019.

\bibitem{Nicodemi01}
Mario Nicodemi and Henrik~Jeldtoft Jensen.
\newblock Aging and memory phenomena in magnetic and transport properties of
  vortex matter: a brief review.
\newblock {\em J. Phys A}, 34:8425, 2001.

\bibitem{Becker14}
Nikolaj Becker, Paolo Sibani, Stefan Boettcher, and Skanda Vivek.
\newblock Temporal and spatial heterogeneity in aging colloids: a mesoscopic
  model.
\newblock {\em J. Phys.: Condens. Matter}, 26:505102, 2014.

\bibitem{Nicholson16}
 A. Nicholson and  P. Sibani.
\newblock Cultural Evolution as a Non-Stationary Stochastic Process.
\newblock {\em Complexity }21:214-223

\bibitem{Arthur17}
R. Arthur, A. Nicholson, P. Sibani and M. Christensen.
\newblock The Tangled Nature Model for organizational ecology.
\newblock {\em Comput Math Organ Theory }23:1-31, 2017.

\bibitem{Alba86}
M.~Alba, M.~Ocio, and J.~Hammann.
\newblock Ageing process and {R}esponse {F}unction in {S}pin {G}lasses: an
  {A}nalysis of the {T}hermoremanent {M}agnetization {D}ecay in
  {A}g:{M}n(2.6{\%}).
\newblock {\em Europhys. Lett.}, 2:45--52, 1986.

\bibitem{Zotev03}
{V.S. Zotev, G.F. Rodriguez, G.G. Kenning, R. Orbach, E. Vincent and J.
  Hammann}.
\newblock {Role of Initial Conditions in Spin-Glass Aging Experiments.}
\newblock {\em Phys. Rev. B}, 67:184422, 2003.

\bibitem{Vincent91}
E.~Vincent.
\newblock Slow dynamics in spin glasses and other complex systems.
\newblock In D.~H. Ryan, editor, {\em Recent progress in random magnets}, pages
  209--246. Mc Gill University, 1991.

\bibitem{Rodriguez03}
G.~F. Rodriguez, G.~G. Kenning, and R.~Orbach.
\newblock {Full Aging in Spin Glasses}.
\newblock {\em Phys. Rev. Lett.}, 91:037203, 2003.

\bibitem{Kenning06}
{G.G. Kenning, G.F. Rodriguez and R. Orbach}.
\newblock End of aging in a complex system.
\newblock {\em Phys. Rev. Lett.}, 97:057201, 2006.

\bibitem{Rieger93}
H.~Rieger.
\newblock Non-equilibrium dynamics and aging in the three dimensional {I}sing
  spin-glass model.
\newblock {\em J. Phys. A}, 26:L615--L621, 1993.

\bibitem{Kisker96}
J.~Kisker, L.~Santen, M.~Schreckenberg, and H.~Rieger.
\newblock Off-equilibrium dynamics in finite-dimensional spin-glass model.
\newblock {\em Phys. Rev. B}, 53:6418--6428, 1996.

\bibitem{Sibani07}
{P. Sibani}.
\newblock {Linear response in aging glassy systems, intermittency and the
  Poisson statistics of record fluctuations}.
\newblock {\em Eur. Phys. J. B}, 58:483--491, 2007.

\bibitem{Belletti09}
F. Belletti, A. Cruz, L.A. Fernandez, A. Gordillo-Guerrero, M. Guidetti, A. Maiorano, F. Mantovani, 
E. Marinari, V. Martin-Mayor, J. Monforte, A. Mu\~{n}oz Sudupe, D. Navarro, G. Parisi, S. Perez-Gaviro, 
J. J. Ruiz-Lorenzo, S. F. Schifano, D. Sciretti, A. Tarancon, R. Tripiccione and D. Yllanes (Janus Collaboration).
\newblock {\em J. Stat. Phys.}, 135:1121, (2009)

\bibitem{BaityJesi17}
M. Baity-Jesi, E. Calore, A. Cruz, L. A. Fernandez, J. M. Gil-Narvion, A. Gordillo-Guerrero, 
D. I\~{n}iguez, A. Maiorano, E. Marinari, V. Martin-Mayor, J. Monforte-Garcia, A. Mu\~{n}oz- Sudupe,
 D. Navarro, G. Parisi, S. Perez-Gaviro, F. Ricci-Tersenghi, J.J. Ruiz-Lorenzo, S. F. Schifano, 
 B. Seoane, A. Tarancon, R. Tripiccione, and D. Yllanes (Janus collaboration).
\newblock {\em Phys. Rev. Lett.}, 118:157202, 2017.

\bibitem{Sibani18}
Paolo Sibani and Stefan Boettcher.
\newblock Mesoscopic real-space structures in spin-glass aging: The
  edwards-anderson model.
\newblock {\em Phys. Rev. B}, 98:054202, 2018.

\bibitem{Edwards75}
S.~F. Edwards and P.~W. Anderson.
\newblock Theory of spin glasses.
\newblock {\em J. Phys. F}, 5:965--974, 1975.

\bibitem{Note1}
We gloss here over the fact that an instantaneous quench is not experimentally
  achievable.

\bibitem{Anderson04}
P.~Anderson, H.~J. Jensen, L.~P. Oliveira, and P.~Sibani.
\newblock Evolution in complex systems.
\newblock {\em Complexity}, 10:49--56, 2004.

\bibitem{Sibani06}
Paolo Sibani.
\newblock Mesoscopic fluctuations and intermittency in aging dynamics.
\newblock {\em Europhys. Lett.}, 73:69--75, 2006.

\bibitem{Sibani13}
Paolo Sibani.
\newblock Coarse-graining complex dynamics:continuous time random walks vs.
  record dynamics.
\newblock {\em EPL}, 101:30004, 2013.

\bibitem{Palmer82}
R.G. Palmer.
\newblock Broken ergodicity.
\newblock {\em Advances in Physics}, 31:669--735, 1982.

\bibitem{Sibani06a}
{P. Sibani, G.F. Rodriguez and G.G. Kenning}.
\newblock Intermittent quakes and record dynamics in the thermoremanent
  magnetization of a spin-glass.
\newblock {\em Phys. Rev. B}, 74:224407, 2006.

\bibitem{Dall01}
Jesper Dall and Paolo Sibani.
\newblock {Faster} {M}onte {C}arlo simulations at low temperatures. {T}he
  waiting time method.
\newblock {\em Comp. Phys. Comm.}, 141:260--267, 2001.

\bibitem{Boettcher18}
Stefan Boettcher, Dominic M. Robe and Paolo Sibani.
\newblock Aging is a log-Poisson process, not a renewal process.
\newblock {\em Phys. Rev. E}, 98:020602, 2018.

\bibitem{Joh99}
Y.~G. Joh, R.~Orbach, G.~G. Wood, J.~Hammann, and E.~Vincent.
\newblock Extraction of the {S}pin {G}lass {C}orrelation {L}ength.
\newblock {\em Phys. Rev. Lett.}, 82:438--441, 1999.

\bibitem{Katzgraber06}
H.~G. Katzgraber, M.~K\"orner, and A.~P. Young.
\newblock Universality in three-dimensional {{Ising}} spin glasses: A {Monte
  Carlo} study.
\newblock {\em Phys. Rev. B}, 73:224432, 2006.

\bibitem{Katzgraber03}
H.~G. Katzgraber and A.~P. Young.
\newblock {Monte Carlo} studies of the one-dimensional {Ising} spin glass with
  power-law interactions.
\newblock {\em Phys. Rev. B}, 67:134410, 2003.

\bibitem{Uhlig95}
 C. Uhlig, K. H. Hoffmann and Paolo Sibani.
\newblock Relaxation in self similar hierarchies.
\newblock {\em Zeitschrift f\"{u}r Physik B}, 96:400--416, 1995.

\bibitem{Sibani90a}
P.~Sibani, J.~M. Pedersen, K.~H. Hoffmann, and P.~Salamon.
\newblock Monte {C}arlo dynamics of optimization problems, a scaling
  description.
\newblock {\em Phys. Rev. A}, 42:7080--7086, 1990.

\bibitem{Sibani94a}
P.~Sibani and J-O. Andersson.
\newblock Excitation morphology in short range {I}sing spin glasses.
\newblock {\em Physica A}, 206:1--12, 1994.

\bibitem{Andersson96}
J.-O. Andersson and P.~Sibani.
\newblock Domain growth and thermal relaxation in spin glasses.
\newblock {\em Physica A}, 229:963--966, 1996.

\bibitem{Wood10}
G.~G. Wood.
\newblock The spin glass correlation length and the crossover from three to
  two dimensions.
  \newblock {\em Journal of Magnetism and Magnetic Materials}, 322:1775--1778,
  2010.
  
  \bibitem{Guchhait17}
Samaresh Guchhait and Raymond L. Orbach.
\newblock Magnetic Field Dependence of Spin Glass Free Energy Barriers.
\newblock {\em Phys. Rev. Lett.} 118:157203, 2017.
  \end{thebibliography}
  \end{document}